\documentclass{elsart}

\usepackage{epsf}

\begin{document}
\begin{frontmatter}

\title{Are Mean-Field Spin-Glass Models Relevant for the Structural 
       Glass Transition?}

\author{A. Crisanti\thanksref{mail1}}
\address{Dipartimento di Fisica, Universit\`a di Roma ``La Sapienza''
         and \\
         Istituto Nazionale Fisica della Materia, Unit\`a di Roma\\
         P.le Aldo Moro 2, I-00185 Roma, Italy}
and
\author{F. Ritort\thanksref{mail2}}
\address{Physics Department, Faculty of Physics \\
         University of Barcelona, Diagonal 647, 08028 Barcelona,  Spain
        }
\thanks[mail1]{e-mail: andrea.crisanti@phys.uniroma1.it}
\thanks[mail2]{e-mail: ritort@ffn.ub.es}

\begin{abstract}
We  analyze the properties of the energy landscape of {\it finite-size} 
fully connected $p$-spin-like models whose high temperature phase is 
described, in the thermodynamic limit, by the schematic Mode Coupling Theory 
of super-cooled liquids.
We show that {\it finite-size} fully connected $p$-spin-like models,
where activated processes are possible, 
do exhibit properties typical of real
super-cooled liquid when both are near the critical glass transition.
Our results support the conclusion that fully-connected  
$p$-spin-like models are the natural statistical mechanical models for 
studying the glass transition in super-cooled liquids. 
\end{abstract}

\begin{keyword}
Glass transition, spin-glass, random models

PACS: 64.70.Pf, 75.10.Nr, 61.20.Gy, 82.20.Wt
\end{keyword}

\end{frontmatter}

\section{Introduction}
In recent years a significant effort has been devoted to the understanding 
of glass-forming systems. Recent theoretical and numerical results
clearly show that the slowing down of the dynamics near the glass transition 
is strongly connected to the potential energy landscape geometry. 
The trajectory of 
the representative point in the configuration space can be viewed as a path 
in a multidimensional potential energy surface \cite{GO}. 
The dynamics is therefore 
strongly influenced by the topography of the potential energy landscape: 
local minima, barriers heights, basins of attraction an other topological 
properties all influence the dynamics. 

The potential energy surface of a super-cooled liquid 
contains a large number of local minima, called {\it inherent structures}
(IS) by Stillinger \cite{S95}. 
All states that under local energy minimization will flow into the same IS
define the {\it basin} of the IS (valley). With this pictures in mind
the time evolution of the system can be seen as the result of two
different processes: thermal relaxation into basins
({\it intra-basin} motion)  and 
thermally activated potential energy barrier crossing between different
basins ({\it inter-basin} motion). 
When the temperature is lowered down to the order of the critical 
Mode Coupling Theory (MCT) temperature $T_{MCT}$ the inter-basin 
motion slows down and the relaxation dynamics is dominated by 
the slow thermally activated crossing of potential energy barriers 
\cite{SSDG99,CR99}.
If the temperature is further reduced the
relaxation time eventually becomes of the same order of the observation 
time and the system falls out of equilibrium since there is not
enough time to cross barriers and equilibrate. 
This define the ``experimental''glass transition temperature $T_g$.
The regime between $T_{MCT}$ and $T_g$ cannot be described by the MCT 
since it neglects activated processes responsible for barrier crossing. 
In MCT the relaxation time diverges at $T_{MCT}$, leading to $T_g=T_{MCT}$, 
and the dynamics
remains confined into a single basin forever. 

The essential features of MCT for glass-forming systems 
are also common to the high temperature phase of some fully connected 
spin glass models \cite{KT87},
the most well known being the 
spherical $p$-spin spin glass model \cite{CS92,CHS93}. 
We shall call these models 
{\it mean-field $p$-spin-like} glass models. 
As a consequence at the critical temperature $T_{MCT}$, called $T_D$ in 
$p$-spin language, an ergodic to non-ergodic transition 
takes place. 
Below this temperature the system is dynamically confined to a metastable
state (a basin) \cite{CS95} since
relaxation to true equilibrium can only take place via 
activated processes, absent in mean-field models.
For these systems, nevertheless, it is known that the true equilibrium 
transition to a low temperature phase occurs below $T_D$ at the static 
critical temperature $T_c$, also denoted by $T_{1rsb}$ \cite{CS92}. 
This is the analogous of the Kauzmann temperature $T_k$ for liquids.
The glass transition temperature $T_g$ of real systems sits somewhere
in between $T_c$ and $T_D$. 
This transition, obviously, cannot be reached even on infinite time in 
mean-field models.

Despite these difficulties mean-field models, having the clear advantage
of being analytically tractable, have been largely 
used to study the properties of fragile glassy systems, 
especially between the dynamical temperature $T_D$ and the static 
temperature $T_c$. 
The picture that emerges is however not complete since
activated process cannot be captured by mean-field models. 
Therefore the relevance of mean-field results for real systems
cannot be considered completely stated. 

Only recently activated processes in mean-field-like models
have been invesigated in 
extended numerical investigation of {\it finite-size} 
fully-connected $p$-spin-like models \cite{CR99,CR99b}.
Comparing the results with the observed behavior of super-cooled liquids 
near $T_{MCT}$ we can conclude that, once activated process are allowed, 
mean-field $p$-spin-like models are highly valuable for a deep 
understanding of the glass transition in real systems.

We report the main results obtained
for the Ising-spin Random
Orthogonal Model (ROM) \cite{MPR94,PP95}, defined by the Hamiltonian
\cite{MPR94,PP95},
%\begin{equation}
%\label{eq:ham}
$
  H = - 2 \sum_{ij} J_{ij}\, \sigma_i\, \sigma_j 
$
%\end{equation}
where $\sigma_i=\pm 1$ are $N$ Ising spin variables, 
and $J_{ij}$ is a $N\times N$ random 
symmetric orthogonal matrix with $J_{ii}=0$.  
For $N\to\infty$ this model has the
same thermodynamic properties of the $p$-spin model:
a dynamical transition at $T_D=0.536$, 
with threshold energy per spin $e_{th} = E_{th}/N = -1.87$, 
and a static transition at
$T_c=0.25$, with critical energy per spin $e_{1rsb}= -1.936$ 
\cite{MPR94,PP95}. 

\section{Thermodynamics of Inherent Structures: How to evaluate the 
         configurational entropy}

The free energy analysis (TAP) \cite{CS95,PP95} reveals 
that the phase space 
is composed by an exponentially large (in $N$) number 
of different basins, separated 
by infinitely large (for $N\to\infty$) barriers. Each basin is
unambiguously labelled by the value of the energy density $e$ 
of the local minimum contained within it, i.e. 
the  IS of the system. 
In this picture
the dynamical transition is associated with IS having the
largest basin of attraction for $N\to\infty$, while the static transition
with IS with the lowest accessible free energy (vanishing configurational
entropy) \cite{KW87,CS95}.

In the mean-field limit, the allowed values of $e$ are between 
$e_{1rsb}$ and $e_{th}$. 
Solutions with $e$ larger than $e_{th}$ are unstable (saddles),
while solutions with $e$ smaller than $e_{1rsb}$ have negligible statistical
weight. Moreover in the $N\to\infty$
limit IS with $e=e_{th}$ attract most 
(exponentially in $N$) of the states and dominate the behavior of the 
system. Other IS are irrelevant for $N\to\infty$.
For finite $N$ the scenario is different since not only the basins of 
IS with $e < e_{th}$ acquire statistical weight,
but it may happen that solutions with $e>e_{th}$ 
and few negative directions (saddles with few downhill directions) 
become stable, simply because there are not 
enough degrees of freedom to hit them. 

%\begin{figure}[hbt]
%\epsfxsize=10truecm\epsfysize=8truecm
%\epsfbox{proc.fig1.eps}
%\caption{Temperature dependence of $\langle e(T)\rangle$ for
%         $N=48$ (square), $N=300$ (circle) and $N=1000$ (triangle). 
%	The average is over $10^3$ different equilibrium
%	configurations at temperature $T$. 
%	 The horizontal line is the $N\to\infty$ limit.
%         The arrows indicate the critical temperatures $T_D$  
%         and $T_c$ (see text). 
%         The dotted line is the curve obtained from the
%         configurational entropy for large $N$.
%         Inset: size dependence of 
%         $\langle e(T=3)\rangle$ as a function $N^{-\alpha}$ 
%         with $\alpha=0.2$ extrapolated down to the $N\to\infty$ 
%         theoretical result $-1.87$ (triangle).
%}
%\label{fig:f1}
%\end{figure}

To get more insight the IS-structure of finite systems we follow 
Stillinger and Weber \cite{SW82} and decompose the
partition sum into a sum over basins of different IS and
a sum within each basin. Collecting all IS with the same energy
$e$, denoting with $\exp [N s_c(E)]\,de$ the number of IS 
with energy between 
$e$ and $e+de$, and shifting the energy of each basin with
that of the associated IS,
the partition sum can be rewritten as \cite{SW82}
\begin{equation}
\label{eq:part}
  Z_N(T)\simeq \int d e \exp\, N\,\left[-\beta e + s_c(e)
                                     -\beta f(\beta,e)
                               \right]
\end{equation}
where $f(\beta,e)$ can be seen as the free energy density of the
system when confined in one of the basin associated with IS
of energy $e$. The function $s_c(e)$ is the {\it configurational
entropy density} also called {\it complexity}. 
 From the partition 
function we can compute the average internal energy density
$u(T) = \langle e +\partial (\beta f) / \partial \beta \rangle =$
$\langle e(T) \rangle + \langle \Delta e(T)\rangle$. The first term is 
the average energy of the IS relevant for the thermodynamics at 
temperature $T$, while the second is the contribution 
from fluctuations inside the associated basins.
In the limit $N\to\infty$ only IS with $e=e_{th}$ contribute 
and $\lim_{N\to\infty} \langle e(T) \rangle = e_{th}$ for any $T>T_D$. 
For finite $N$, and $T$ not too close to
$T_D$, the thermodynamics is dominated by IS with $e>e_{th}$ and
$\langle e(T)\rangle > e_{th}$ \cite{CR99}. This is more 
evident from the (equilibrium) probability distribution of $e$
since it is centered about $\langle e(T)\rangle$ indicating that IS
with $e\simeq \langle e(T)\rangle$ have the largest basins.
This scenario has been also observed in real 
glass-forming systems\cite{OTW88,SDS98,SST98,Parisi,KST99}.

\begin{figure}[hbt]
\epsfxsize=10cm\epsfysize=8cm
\epsfbox{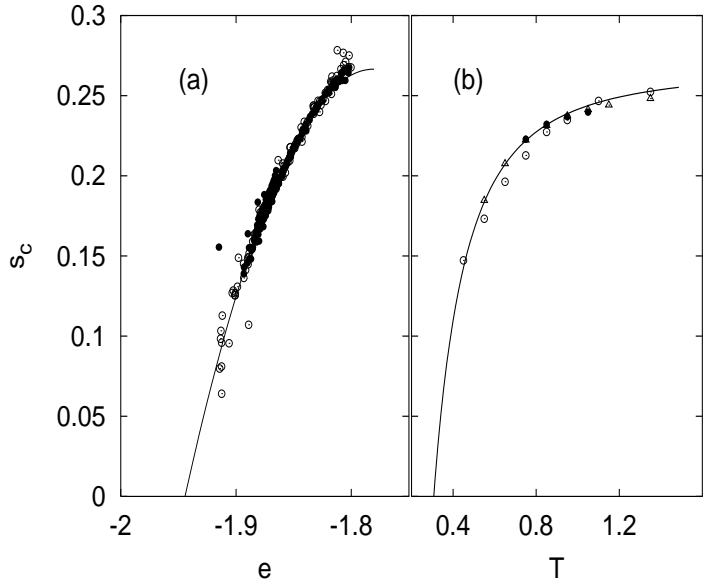}
\caption{(a) Configurational entropy as a function of energy.
         The data are from system sizes $N=48$ (empty circle) and $N=300$
         (filled circle), and temperatures $T=0.4$, $0.5$, $0.6$, $0.7$, 
         $0.8$, $0.9$ and $1.0$. For each curve the unknown constant has
         been fixed to maximize the overlap between the data and
         the theoretical result \protect\cite{PP95}. The line is the
	 quadratic best-fit.
         (b) Configurational entropy density as a function of temperature.
         The line is the result from the best-fit of $s_c(e)$ 
         while the symbols are the results from the temperature integration
         of 
         $d\, s_c(T) / d\, \langle e(T)\rangle = T^{-1}$
         for $N=48$ (empty circle), 
         $N=300$ (empty triangle) and $N=1000$ (filled circle). 
}
\label{fig:f2}
\end{figure}

 From the knowledge of IS-energy distribution we can reconstruct the
complexity $s_c(e)$ since  from eq. (\ref{eq:part}) 
the probability that an equilibrium
configuration at temperature $T=1/\beta$ lies in a basin associated with
IS of energy between $e$ and $e+de$ is:
%\begin{equation}
%\label{eq:prob}
$
 P_N(e,T) =  \exp\, N\,\left[-\beta e + s_c(e)
                                     -\beta f(\beta,e)
                               \right] / Z_N(T).
$
%\end{equation}
In the temperature range where this applies, the
curves $\ln P_N(e,T) + \beta e$ are equal, except for a temperature
dependent factor $\ln Z_N(T)$, to $s_c(e)-\beta f(\beta,e)$. If 
the $e$-dependence of $f(\beta,e)$ can be neglected, 
then it is possible to
superimpose the curves for different temperatures, see Fig.
\ref{fig:f2} (a). 
The data collapse is rather good for $e < -1.8$. 
Above the curves cannot be superimposed anymore indicating that the
$e$-dependence of $f(\beta,e)$ cannot be neglected.  In
liquid this is called the anharmonic threshold \cite{SKT99,BH99}.

Direct consequence of $f(\beta,e) \simeq f(\beta)$ for $e<-1.8$ is that
in this range the partition function can be written as the product of an
intra-basin contribution [$\exp(-N\beta f$)] 
and of a configurational contribution which
depends only on the IS energy densities distribution.
The system can then be considered as composed by two independent subsystems:
the intra-basin subsystem describing the equilibrium when confined within 
basins, and the IS subsystem describing equilibrium via activated processes
between different basins.
As the temperature is lowered
and/or $N$ increased the two processes get more separated in time and the 
separation becomes more and more accurate. 
A scenario typical of super-cooled liquids near the MCT 
transition \cite{ST97,SSDG99}.

\begin{figure}[hbt]
\epsfxsize=10cm\epsfysize=8cm
\epsfbox{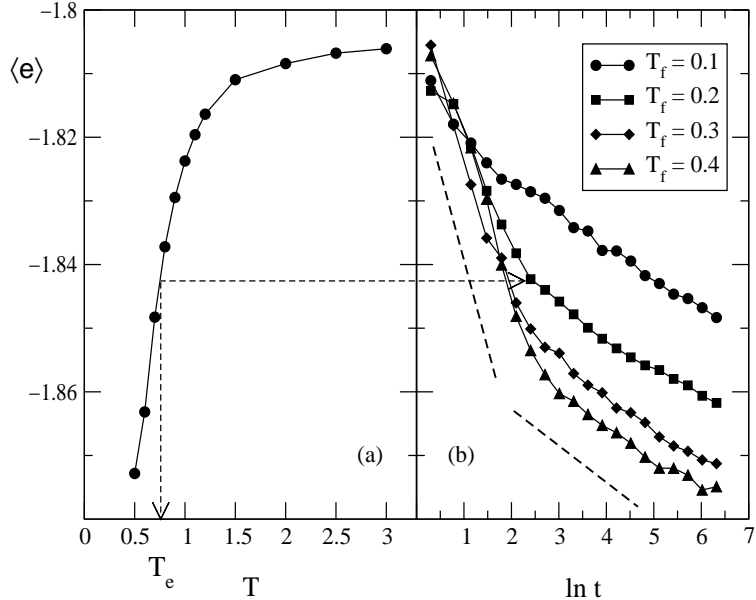}
\caption{Average inherent structure energy in equilibrium as a function
         of temperature $(a)$ and as a function of time during 
         the non-equilibrium process $(b)$. The system size is $N=300$,  
         $T_i= 3.0$ and (top to bottom)
         $T_f= 0.1$, $0.2$, $0.3$ and $0.4$ (panel b). 
}
\label{fig:e-aging}
\end{figure}

\section{Non-equilibrium behavior: the role of activated processes}

More informations on the IS structure can be obtained from
non-equilibrium relaxation processes.  To study the non-equilibrium
dynamics we quench at time zero the system from an initial equilibrium
configuration at temperature $T_i>T_g$ to a final temperature
$T_f<T_g$ and study the evolution of the average IS energy per spin
$\langle e(t)\rangle$ as function of time, see Fig.\ref{fig:e-aging} (b).  
Two different relaxation processes are clear seen.
A first regime
independent of $T_f$, and a second regime independent of 
both $T_i$ and $T_f$. The final temperature $T_f$ controls the cross-over 
between the two regimes. A similar behavior has been observed in 
molecular dynamics simulations of super-cooled liquids \cite{KST99}. 
The two regimes are associated with different relaxation processes.
In the first part the system has enough energy and
relaxation is mainly due to {\it path search} out of basins through
saddles of energy lower than $k_{B}T_f$.
This part depends only on the initial 
equilibrium temperature $T_i$ since it sets the initial phase space region.
Different $T_i$ leads to different power law. 
In particular relaxation must slow down as 
$T_i$ decreases since we expect that lower states are surrounded by higher 
barriers, in agreement with numerical data \cite{CR99b}. 

During this process the system explores deeper and deeper valleys (basins)
while decreasing its energy. The process stops when all barrier heights
become of $O(k_{B}T_f)$. From now on the relaxation proceeds only via 
activated process. 
A first consequence is that lower the final temperature $T_f$ 
shorter the first relaxation, in agreement with our findings [See figures
\ref{fig:e-aging}].

The analysis of 
the distance between the instantaneous system state 
and the corresponding IS, counting  the number of single spin flip 
needed to reach the IS, reveals that 
for all times the systems stays in configurations 
few spin flips away from an IS. A similar study starting from 
equilibrium configurations at temperature 
$T_e(\langle e(t)\rangle)$ evaluated comparing panels (a)
and (b) of figure \ref{fig:e-aging} \cite{KST99} leads to similar numbers.
We then conclude that during relaxation the aging system explores 
the same type of minima (and basins) visited in equilibrium
at temperature $T_e$. 
Direct consequence is that 
once the system has reached the activated regime  there cannot be
memory of the initial $T_i$, and all curves with different $T_i$ but
same $T_f$ should collapse for large time \cite{CR99b}.

\section{Conclusions}

To summarize, we have shown that {\it finite-size}
mean-field $p$-spin-like models are good candidates for studying the glass
transition. The key point is that near the glass
transition the thermodynamics of the systems is dominated by the IS
distributions, therefore all systems with similar IS distributions
should have similar behavior. 
Finite-size mean-field $p$-spin-like models have the double advantage 
of being analytically tractable for $N\to\infty$ and easily simulated 
numerically for finite $N$, offering good models to analyze the glass 
transition.

%\begin{ack}
%We thank for useful discussions 
%C. Donati, U. Marini Bettolo, F. Sciortino and P. Tartaglia.
%\end{ack}

%%%%%%%%%%%%%%%%%%%%%%%%%% REFERENCES %%%%%%%%%%%%%%%%%%%%%


\begin{thebibliography}{99}

\bibitem{GO} M. Goldstein, Phys. Rev. {\bf 51}, 3728 (1969)

\bibitem{S95}
	F. H. Stillinger,
	Science {\bf 267}, 1935 (1995)

\bibitem{SSDG99}
	T. B. Schr{\o}der, S. Sastry, J. C. Dyre and S. C. Glotzer,
	cond-mat/9901271

\bibitem{CR99}
	A. Crisanti and F. Ritort, cond-mat/9907499
	
\bibitem{KT87}
	T. R. Kirkpatrick and D. Thirumalai, 
	Phys. Rev. Lett. {\bf 58}, 2091 (1987)

\bibitem{CS92}
	A. Crisanti, and H. J. Sommers,
	Z. Physik B{\bf 87}, 341 (1992)

\bibitem{CHS93}
	A. Crisanti, H. Horner and H. J. Sommers,
	Z. Physik B{\bf 92}, 257 (1993)

\bibitem{CS95}
	A. Crisanti and H. J. Sommers,
        J. Phys. I France {\bf 5}, 805 (1995)

\bibitem{CR99b}
	A. Crisanti and F. Ritort,
        in preparation (1999).

\bibitem{MPR94}
	E. Marinari, G. Parisi and F. Ritort,
	J. Phys. A (Math. Gen.) {\bf 27}, 7847 (1994).

\bibitem{PP95}
	G. Parisi and M. Potters, J. Phys. A (Math. Gen.) {\bf 28}, 5267 (1995)


\bibitem{KW87}
	T.R. Kirkpatrick and P.G. Wolynes,
	Phys. Rev. B{\bf 36}, 8552 (1987).

\bibitem{SW82}
	F. H. Stillinger and T. A. Weber,
	Phys. Rev. A{\bf 25}, 978 (1982)

\bibitem{OTW88}
	I. Ohmine, H. Tanaka and P.G. Wolynes,
	J. Chem Phys. {\bf 89}, 5852 (1998);
        H. Tanaka,
	Nature {\bf 380}, 328 (1996)

\bibitem{SDS98}
	S. Sastry, P.G. Debenedetti and F. H. Stillinger,
	Nature {\bf 393}, 554 (1998)

\bibitem{SST98}
	F. Sciortino, S. Sastry  and P. Tartaglia,
	cond-mat/9805040

\bibitem{Parisi}
         B. Coluzzi, M. Mezard, G. Parisi and P. Verrocchio,
         cond-mat/9903129
\bibitem{KST99}
	W. Kob, F. Sciortino  and P. Tartaglia,
	cond-mat/9905090

\bibitem{SKT99}
	F. Sciortino W. Kob and P. Tartaglia,
	cond-mat/9906081

\bibitem{BH99}
	S. B\"{u}chner and A. Heuer,
	cond-mat/9906280

\bibitem{ST97}
	F. Sciortino and P. Tartaglia,
        Phys. Rev. Lett. {\bf 78}, 2385 (1997)

\end{thebibliography}
\end{document}